\documentclass[authoryear]{elsarticle}
\usepackage{latexsym}
\usepackage{float}
\usepackage[latin2]{inputenc}
\usepackage{graphicx}
\usepackage{ae}
\usepackage{aecompl}
\usepackage{setspace}
\usepackage[hidelinks]{hyperref}

\makeatletter
\def\ps@pprintTitle{%
	\let\@oddhead\@empty
	\let\@evenhead\@empty
	\def\@oddfoot{\centerline{\thepage}}%
	\let\@evenfoot\@oddfoot}
\makeatother

		\title{The Budapest Amyloid Predictor and its Applications}
		
	\author[p]{László Keresztes\corref{cor2}}
	\ead{keresztes@pitgroup.org}
	\author[p]{Evelin Szögi\corref{cor2}}
	\ead{szogi@pitgroup.org}
	\author[p]{Bálint Varga}
	\ead{balorkany@pitgroup.org}
	\author[s]{Viktor Farkas}
	\ead{farkasv@caesar.elte.hu}
	\author[s,t]{András Perczel}
	\ead{perczel@chem.elte.hu}
	\author[p,u]{Vince Grolmusz\corref{cor1}}
	\ead{grolmusz@pitgroup.org}
	\cortext[cor1]{Corresponding author}
	\cortext[cor2]{Joint first authors}
	\address[p]{PIT Bioinformatics Group, Eötvös University, H-1117 Budapest, Hungary}
	\address[u]{Uratim Ltd., H-1118 Budapest, Hungary}
	\address[s]{MTA-ELTE Protein Modeling Research Group, Eötvös University, H-1117 Budapest, Hungary}
	\address[t]{Laboratory of Structural Chemistry and Biology, Eötvös University, H-1117, Budapest, Hungary}
	
\begin{document}

		\date{}

\begin{abstract}
	\noindent{\bf Summary:} The amyloid state of proteins is widely studied with relevancy in neurology, biochemistry, and biotechnology. In contrast with amorphous aggregation, the amyloid state has a well-defined structure, consisting of parallel and anti-parallel $\beta$-sheets in a periodically repeated formation. The understanding of the amyloid state is growing with the development of novel molecular imaging tools, like cryogenic electron microscopy.  Sequence-based amyloid predictors were developed by using mostly artificial neural networks (ANNs) as the underlying computational techniques. From a good neural network-based predictor, it is a very difficult task to identify those attributes of the input amino acid sequence, which implied the decision of the network. Here we present a Support Vector Machine (SVM)-based predictor for hexapeptides with correctness higher than 84\%, i.e., it is at least as good as the published ANN-based tools. Unlike the artificial neural networks, the decision of the SVMs are much easier to analyze, and from a good predictor, we can infer rich biochemical knowledge. 	
	
	\noindent{\bf Availability and Implementation:} The Budapest Amyloid Predictor webserver is freely available at \url{https://pitgroup.org/bap}. 
	
	\noindent{\bf Contact:} grolmusz@pitgroup.org 	
\end{abstract}
	
	\maketitle	

\section{Introduction and motivation} 

The primary structure of the proteins is characterized by their amino acid sequence. While the primary structure determines the spatial folding of the proteins, and, consequently, all chemical and biological properties of the given protein, inferring those properties from the amino acid sequence is a very difficult task. Here we consider the amyloid predictors: tools, which tell us if a given amino acid sequence has or has not the propensity to become amyloid.

Amyloids are misfolded protein aggregates \citep{Horvath2019,Taricska2020}, which -- in contrast with the unstructured aggregates -- have a well-defined structure, comprising parallel $\beta$-sheets \citep{Takacs2019,Takacs2020}. Amyloids are present in numerous organisms in biology: for example, in healthy human pituitary secretory granules \citep{Maji2009}, in the immune system of certain insects \citep{Falabella2012}, the silkmoth chorion and some fish choria \citep{Iconomidou2008}, in human amyloidoses and several neuro-degenerative diseases \citep{Soto2006}. 

Most recently, on the analogy of the naturally occurring anti-herpes activity of $\beta$-amyloids, synthetic amyloid peptides were developed, acting as amyloidogenic aggregation cores in certain viral proteins with high specificity \citep{Michiels2020}. This way, new amyloid-based antiviral pharmaceuticals can be developed in the very near future: the specific aggregation cores turn the viral proteins into insoluble amyloids. Consequently, potential amyloidogenecity may have direct pharmaceutical relevance.

Sequence-based amyloid predictors would help the understanding and the exploitation of the amyloid state of the proteins: instead of the difficult, costly, and slow wet-laboratory tests, we can use the predictor on thousands or millions of inputs for enlightening the amyloidogenecity of the proteins. A very recent review \citep{Santos2020} covers the sequence-based  amyloid-predictors, applying different strategies like AGGRESCAN \citep{Conchillo-Sole2007}, Zyggregator \citep{Tartaglia2008a}, netCSSP \citep{Kim2009a} and APPNN \citep{Familia2015}, among others. 

In the last several years, the six amino acid long peptides have become a model of studying amyloid formation \citep{Beerten2015, Louros2020, Louros2020a}. The reason for this is twofold: first: numerous evidence shows the biological relevance of amyloid-forming hexapeptides \citep{Hauser2011,Tenidis2000,Reches2004,Iconomidou2006,Beerten2015}; and second:  one can form $20^6=$ 64 million hexapeptides from the 20 amino acids, which is a large -- but not too large -- and rich space of model molecules, whose structures are less complex and, therefore, easier to be dealt with as larger model spaces.  

The APPNN predictor applies a machine-learning approach by training on 296 hexapeptides, selected from various sources, then predicts if a given hexapeptide is amyloidogenic or not. For longer sequences, it screens six amino-acid long sliding windows in longer polypeptide-chains to predict if they would form amyloid structures.   

In this contribution, we present a Support Vector Machine (SVM) predictor for hexapeptides, with better accuracy (84\%) than most of the neural network-based tools (see \citep{Familia2015} for a tabular comparison of the accuracy of those tools). The main advantage of our new predictor is its (i) simplicity, (ii) free on-line availability, and (iii) easy applicability for inferring location-dependent amyloidogenic properties of amino acids, as we describe below. 

We note that neural network-based predictors are neither simple nor easy-to-apply, and inferring the causality of their classifications is a very difficult task.

\section{Materials and Methods}

For the construction of the Budapest Amyloid Predictor, we have applied an artificial intelligence tool, the Support Vector Machine architecture \citep{Cortes1995}. In Support Vector Machines, $n+m$ data points are corresponded to $n+m$ vectors, each of $k$ dimensions, $x_1,x_2,\ldots,x_n$ and $y_1,y_2,\ldots,y_m$, and the goal is to find a hyperplane which optimally separates the $x$ and the $y$ datapoints. Usually, the dataset is partitioned into a training and a testing subset: the first one is applied in the construction of the SVM, the second one is used for testing.

\begin{figure}[t!]
	\begin{center}
		\includegraphics[width=8cm]{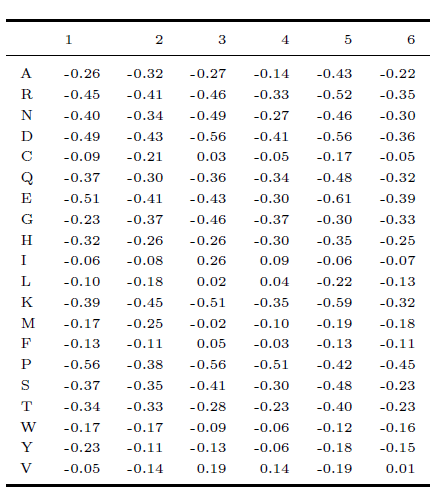}
		\centerline{Table 1 }
	\end{center}
\end{figure}

We have used the Waltz database \citep{Beerten2015, Louros2020} of 1415 hexapeptides, annotated to be amyloidogenic (514 peptides) or not-amyloidogenic (901 peptides). The annotation was made by Thioflavin-T binding assays and literature search \citep{Beerten2015, Louros2020}; consequently, it is based on experimental evidence. Similarly, as in \citep{Familia2015}, two vectorial representations of the hexapeptides were considered: The first is the simple translation of the 20 amino acid names into vectors: each amino acid was corresponded to a length-20 0-1 vector, with a single 1-coordinate identifying the amino acid (called orthogonal representation). This way, a hexapeptide is described by a 120-dimensional 0-1 vector. 

The second one is based on a physico-chemical property database of 553 properties AAindex \citep{Kawashima2008}; \url{https://www.genome.jp/aaindex/}. In this representation, each amino acid  corresponds to a 553-dimensional vector, a hexapeptide to a 6 x 553 = 3318-dimensional vector.

From the 1415 (514 amyloids, 901 non-amyloids) hexapeptides found in Waltz database, we selected 158 amyloid and 309 non-amyloid hexapeptides randomly for the test set (roughly 33\% ). We used the remaining hexapeptides for training our linear SVM. We used the sklearn LinearSVC object from the SciKit-learn Python library \citep{scikit-learn} for constructing the classifier. 

The orthogonal representation yielded an approximately 80 \% accuracy, while the AAindex-based a much better accuracy; because of this, we have chosen the second, AAindex-based representation in what follows.

The classifier simply computes the sign of the $w\cdot z+b$ values for the 3318-long $z$ vectors, corresponding to a hexapeptide, where $w$ is a 3318-dimensional weight vector, and $b$ is a scalar, and if this sign is positive, then the prediction is ``amyloidogenic'', otherwise it is ``non-amyloidogenic''. 

On the 467 test examples, we achieved 127 true positives, 31 false positives, 266 true negatives, 43 false negatives. The resulting classifier's performance for unseen examples is 0.8415 $\pm$ 0.0331 with 95\% confidence. Based on test performance: ACC = 0.84, TPR=0.75, TNR=0.9, PPV=0.8, NPV=0.86, (that is, accuracy, true positive ratio, true negative ratio, positive predictive value, negative predictive value, resp.). The accuracy of our SVM is better or on par with that of APPNN \cite{Familia2015}.

\section{Implementation and Usage}

The Budapest Amyloid Predictor webserver is available at the site \url{https://pitgroup.org/bap/}. The user needs to input a hexapeptide with 6 capital letters, and the server returns the prediction for the query, plus the predictions of all 114 (= 6 x 19) 1-Hamming-distance neighbors of the query. If the hexapeptide is listed in the Waltz DB, then the ``known`` word appears next to prediction; otherwise, the ``predicted'' word. 

\begin{figure}[t!]
	\begin{center}
		\includegraphics[width=8cm]{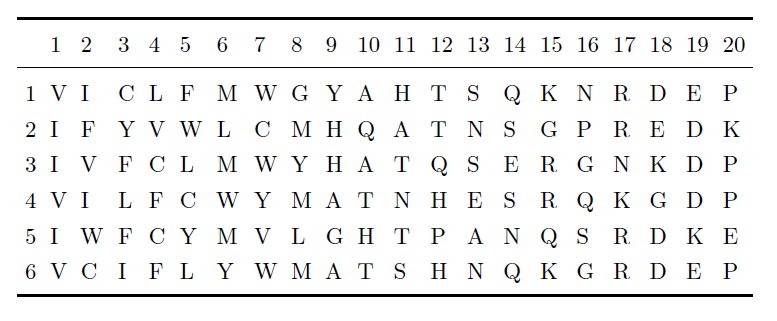}
		\centerline{Table 2 }
	\end{center}
\end{figure}

\subsection*{The Amyloid Effect Matrix}

One of the greatest advantages of the SVM prediction is that we can easily see the reasons behind the decision of the model. The following matrix enlightens the details of the decision of the SVM. Clearly, by representing every amino acid by a 553-dimensional vector is highly redundant, since we have only 20 amino acids: that is, only 20 different 553-dimensional vector exists in this representation. Therefore, we can write with $\ell=553$:
\vskip -0.3cm
$$
w\cdot z=\sum_{i=1}^{6\ell}w_iz_i=\sum_{j=1}^6  \ \ \sum_{i=(j-1)\ell+1}^{j\ell}w_iz_i \eqno{(1)}
$$

For each fixed $j=1,2,\ldots,6$ the $\ell=553$ $z_i'$s are determined by the $j^{th}$ amino acid of the hexapeptide, and this way, all the possible $6 x 20=120$ second sums (for six positions and 20 amino acids) can be pre-computed. Table 1 lists these pre-computed values, the 6 values of $j$ correspond to the columns, the amino acids to the rows:

Clearly, the value of (1) can now be computed by adding exactly one item from each column, determined by the first, second,...,sixth amino acid of the hexapeptide, plus the value of $b=1.083$. For example, one can easily classify the hexapeptide AAEEAA by computing the sign of $(-0.26 - 0.32 -0.43 -0.30 -0.43 -0.22 + 1.083)=-0.88$, that is, -1, which predicts that AAEEAA is {\bf not} amyloidogenic. 

By observing Table 1, one can easily derive an amyloidogenecity order of the amino acids for each position from 1 through 6: we should just sort the columns in increasing order and substitute the amino acids as follows:

In Table 2, the amyloidogenecity order decreases from left to right. Naturally, proline, the "structure breaker" appears mostly at the right end, but not in every row: in row 5 it is in position 12. This shows a remarkable difference in the amyloidogenecity order of the six positions of the hexapeptides.  	

\section*{Funding}
BV, and VG were partially supported by the VEKOP-2.3.2-16-2017-00014 program, supported by the European Union and the State of Hungary, co-financed by the European Regional Development Fund, LK, ES, and VG by the  NKFI-127909
LK, ES, and VG were supported in part by the EFOP-3.6.3-VEKOP-16-2017-00002 grant, supported by the European Union, co-financed by the European Social Fund. LK, ES AP, VF and VG were partially supported by the ELTE Thematic Excellence Programme (Szint+) supported by the Hungarian Ministry for Innovation and Technology.




\end{document}